\documentclass[aip,graphicx,amsmath,amssymb,amsfonts]{revtex4-1}

\usepackage{bm}
\usepackage{physics}
\usepackage[T1]{fontenc}
\usepackage{mathdots}
\usepackage{graphicx}
\usepackage{subfigure}
\usepackage{adjustbox}
\usepackage{bm}
\usepackage{bbold}
\usepackage{color}
\usepackage{braket}
\usepackage{standalone}
\usepackage{multirow}
\usepackage{tikz}
\usepackage{mathrsfs}
\usepackage{dsfont}
\usepackage{float}
\usepackage{placeins}
\usepackage{comment}

\begin{document}


\title{Essential implications of similarities in non-Hermitian systems} 

\author{Anton Montag}
\email[]{anton.montag@mpl.mpg.de}
\affiliation{Max Planck Institute for the Science of Light, 91058 Erlangen, Germany}
\affiliation{Department of Physics, Friedrich-Alexander-Universit\"at Erlangen-N\"urnberg, 91058 Erlangen, Germany}

\author{Flore K. Kunst}
\email[]{flore.kunst@mpl.mpg.de}
\affiliation{Max Planck Institute for the Science of Light, 91058 Erlangen, Germany}

\date{\today}

\begin{abstract}
In this paper, we show that three different generalized similarities enclose all unitary and anti-unitary symmetries that induce exceptional points in lower-dimensional non-Hermitian systems.
We prove that the generalized similarity conditions result in a larger class of systems than any class defined by a unitary or anti-unitary symmetry.
Further we highlight that the similarities enforce spectral symmetry on the Hamiltonian resulting in a reduction of the codimension of exceptional points.
As a consequence we show that the similarities drive the emergence of exceptional points in lower dimensions without the more restrictive need for a unitary and/or anti-unitary symmetry.
\end{abstract}


\maketitle

\section{Introduction}\label{sec:introduction}

In recent years non-Hermitian (NH) Hamiltonians have attracted increasing attention, and one active branch of research focuses on the role of symmetries in NH systems. \cite{Bergholtz2021}
A complete classification in terms of 38 symmetry classes was derived by Kawabata \textit{et al},\cite{Kawabata2019}
and the topological features of these classes as well as the connection between some of them is studied in the literature. \cite{Kawabata2019,Kawabata2019a}
Further it is generally recognized that certain unitary and anti-unitary symmetries lower the codimension of exceptional points (EPs), which are degeneracies where the eigenvalues and the corresponding eigenvectors coalesce. \cite{Kawabata2019,KatoBook,Heiss2012,Miri2019,Ashida2020}
The generic appearance of EP$n$s, where $n$ is the order of the EP set by the number of coalescing eigenvectors, is determined by the codimension of the EP.
As such unitary and anti-unitary symmetries, which are local in parameter space---namely, parity-time ($\mathcal{PT}$) and anti-$\mathcal{PT}$ symmetry, pseudo-Hermitian symmetry, as well as sublattice symmetry, chiral symmetry and parity-particle-hole ($\mathcal{CP}$) symmetry---inflict symmetries on the spectrum, and therefore reduce the codimension of the EPs. \cite{Sayyad2022,Budich2019,Delplace2021,Yoshida2019,Okugawa2019}

\begin{figure}
    \centering
    \includegraphics[width=\textwidth]{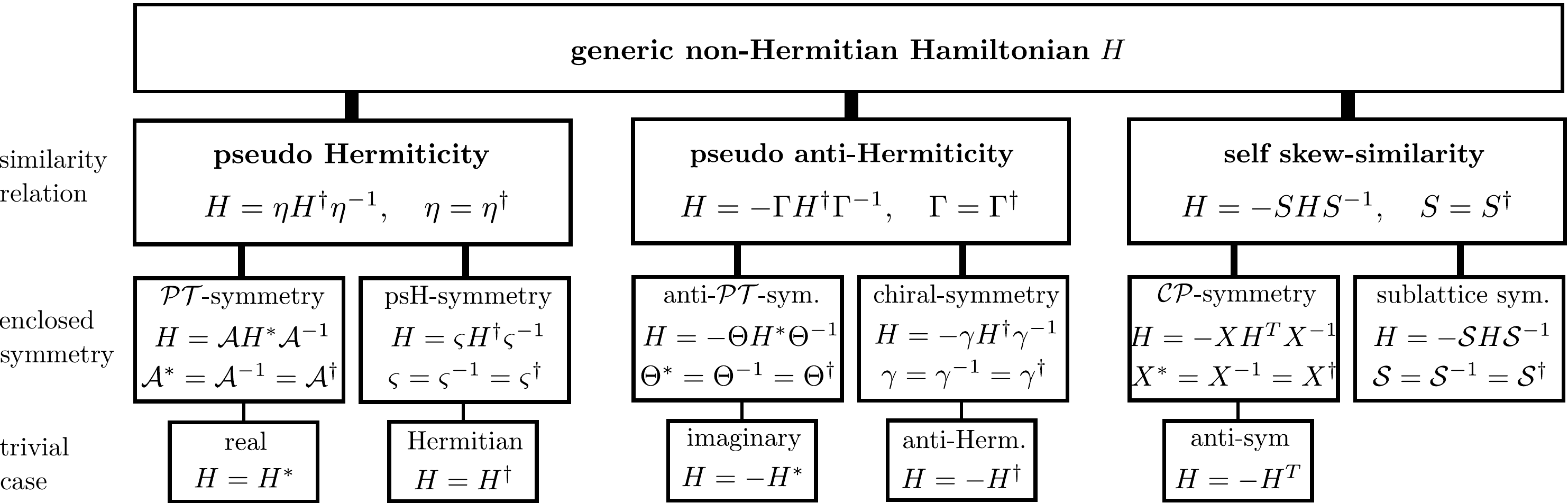}
    \caption{Pairing of all EP-inducing symmetries as special cases of generalized similarities. The three generalized similarity relations that lower the codimension of EPs are given. The different (anti-)unitary symmetries enclosed by the generalized similarities are shown. Trivial cases of the symmetries, where the generator is the identity, are also included in the overview.}
    \label{fig:results}
\end{figure}

The aforementioned symmetries follow from Wigner's theorem \cite{WignerBook} for non-interacting, second-quantized fermionic Hamiltonians $\hat{H}= \sum_{ij} \hat{\Psi}_i^\dagger H_{ij} \hat{\Psi}$ with $\{\hat{\Psi}_i\}$ a set of fermionic annihilation operators, and $H$ the first-quantized Hamiltonian, which says that if $\hat{\mathcal{O}}$ is a symmetry of the system then $[\hat{H},\hat{\mathcal{O}}] = 0$.\cite{Altland1997,Kawabata2019}
For time-reversal $\mathcal{T}$, charge-conjugation $\mathcal{C}$ and parity $\mathcal{P}$ Wigner's theorem can be applied to find symmetry relations of the first-quantized Hamiltonian H. \cite{Altland1997,Kawabata2019}
For first-quantized Bloch Hamiltonians these relations are given by
\begin{align}
    &[\hat{H},\hat{\mathcal{T}}]=0 \quad\implies\quad \mathcal{T} H^*(\bm{k}) \mathcal{T}^{-1} = H(-\bm{k}), \quad \text{and} \\
    &[\hat{H},\hat{\mathcal{C}}]=0 
   \quad \implies\quad \mathcal{C} H^T(\bm{k}) \mathcal{C}^{-1} = -H(-\bm{k}), \quad \text{and} \\
    &[\hat{H},\hat{\mathcal{P}}]=0 \quad\implies\quad \mathcal{P} H(\bm{k}) \mathcal{P}^{-1} = H(-\bm{k}) \, .
\end{align}
The combination of two of these symmetries, namely $\mathcal{PT}$ symmetry, $\mathcal{CP}$ symmetry and chiral symmetry here defined as $\hat{\gamma}=\hat{\mathcal{T}}\hat{\mathcal{C}}$ with the symmetry relation $\gamma H^\dagger \gamma^{-1}=-H$, are local in momentum space and follow directly from Wigner's theorem.
In the Hermitian case this coincides with sublattice symmetry $\hat{\mathcal{S}}$ defined by $\mathcal{S} H \mathcal{S}^{-1}=-H$.
Sublattice symmetry is a linear anti-symmetry that is different from chiral symmetry for non-Hermitian systems.
By expanding the symmetries from Wigner's theorem by sublattice symmetry we find all the symmetries in Fig.~\ref{fig:results} as combinations of different symmetries. In particular, pseudo-Hermitian symmetry is given by a combination of chiral and sublattice symmetry, whereas anti-$\mathcal{PT}$ symmetry comes about through a combination of $\mathcal{PT}$ and sublattice symmetry.
Lastly, we note that these symmetries are consistent with the symmetry classification of random non-Hermitian matrices by Bernard and LeClair, and Kawabata \textit{et al}. \cite{Bernard2002,Kawabata2019}

The (anti-)unitary symmetries can be grouped into three pairs based on the type of constraint they enforce on the set of eigenenergies. \cite{Sayyad2022,Montag2024}
In this work, we show that the spectral symmetries already come about in the presence of similarity relations and not just in the presence of more restrictive symmetries.
These similarity relations, namely pseudo Hermiticity, pseudo anti-Hermiticity and self skew-similarity, naturally pair the anti-unitary and unitary symmetries, cf. Fig.~\ref{fig:results}, and enforce the spectral symmetry.
The symmetries appear as special cases of these three EP-inducing generalized similarities.

The relation of $\mathcal{PT}$ symmetry and pseudo Hermiticity, which denotes the similarity of $H$ and its adjoint $H^\dagger$, is well established.
Quantum mechanics formulated on the basis of $\mathcal{PT}$-symmetric operators was investigated by Bender \textit{et al.}, \cite{Bender1998,Bender2002,BenderBook} and has been related to pseudo-Hermitian operators.
For diagonalizable $\mathcal{PT}$-symmetric operators Mostafazadeh proved pseudo Hermiticity by explicitly showing the similarity between $H$ and $H^\dagger$. \cite{Mostafazadeh12002,Mostafazadeh22002,Mostafazadeh32002,Mostafazadeh2010}
Later this was extended to any finite $\mathcal{PT}$-symmetric Hamiltonian by Zhang \textit{et al.} \cite{Zhang2020}
In Section~\ref{sec:psH} we summarize and expand upon their results by showing a further connection to Hermitian and pseudo-Hermitian symmetric matrices, where we note the subtle difference between pseudo Hermiticity and pseudo-Hermitian symmetry.
Pseudo Hermiticity alone already enforces symmetries on the spectrum, and thus lowers the codimension of EPs, while $\mathcal{PT}$ symmetry and pseudo-Hermitian symmetry constitute two special cases.
Further we include Hermitian Hamiltonians as a special case of pseudo-Hermitian symmetric systems, which has additional spectral symmetry that naturally prevents EPs from emerging.
We also comment on real Hamiltonians as a special case of $\mathcal{PT}$-symmetry.

We find a similar structure for anti-$\mathcal{PT}$-symmetric and chiral-symmetric systems.
Both symmetries enforce pseudo anti-Hermiticity on the systems, which we define in Section~\ref{sec:chiral}.
We prove that all anti-$\mathcal{PT}$-symmetric and chiral-symmetric systems are pseudo anti-Hermitian.
We show the spectral constraint follows from pseudo anti-Hermiticity, and relate anti-Hermiticity to pseudo anti-Hermiticity.
We compare pseudo Hermiticity and pseudo anti-Hermiticity, which have a resembling effect on the spectrum, and point out the similarities and differences between them.

In Section~\ref{sec:skew} we follow the same approach for $\mathcal{CP}$-symmetry and sublattice symmetry, where we find that in both cases the Hamiltonian exhibits self skew-similarity.
This self skew-similarity is the origin of the spectral symmetry.
We note that self skew-similarity behaves differently from pseudo-Hermiticity and pseudo anti-Hermiticity, because it does not relate the Hamiltonian to its adjoint, but instead is a property of the Hamiltonian itself.
This results in differences in the treatment of self skew-similarity.

We provide a conclusion in Section~\ref{sec:conclusion}.

\section{Pseudo-Hermitian systems}\label{sec:psH}

We start from the definition of pseudo-Hermiticity.
A Hamiltonian is called pseudo-Hermitian if there exists an invertible Hermitian matrix $\eta$ such that
\begin{equation}\label{eq:def_psH}
    H=\eta H^\dagger\eta^{-1} ,
\end{equation}
where $H^\dagger$ denotes the conjugate transpose of $H$.

\textbf{Theorem 2.1.} \textit{For a matrix $H\in\mathbb{C} ^{n\times n}$, H is pseudo-Hermitian if and only if there exist a nonsingular Hermitian matrix $\eta$ and a Hermitian matrix $A$ such that $H = \eta A$.}

\textit{Proof.} Insert $H=\eta A$ in the definition of pseudo-Hermiticity using $\eta$ as the similarity matrix.

We give this theorem here to provide a general decomposition of pseudo-Hermitian matrices, and to highlight a method of generating generic pseudo-Hermitian matrices.

For a pseudo-Hermitian matrix $H$ with the eigenstate $\ket{\psi}$, defined by $H\ket{\psi}=\epsilon\ket{\psi}$, it follows from Eq.~(\ref{eq:def_psH}) that
\begin{equation}
    \eta^{-1} H \ket{\psi} = \epsilon \left(\eta^{-1} \ket{\psi}\right) = H^\dagger \left(\eta^{-1}  \ket{\psi}\right) \, ,
\end{equation}
thus $\eta^{-1} \ket{\psi}$ is an eigenstate of $H^\dagger$ with the eigenvalue $\epsilon$, and the eigenvalues of $H$ are either real or appear in complex conjugate pairs, i.e., $\{\epsilon\} = \{\epsilon^*\}$.

\textbf{Theorem 2.2.} \textit{For a matrix $H\in\mathbb{C} ^{n\times n}$, H is pseudo-Hermitian if and only if it is similar to its complex conjugate $H^*$.}

The proof of Theorem 1.2 can be found in Ref.~\citenum{Zhang2020}.
They show that the necessity follows from the definition of pseudo Hermiticity and the similarity of every matrix to its transpose.
The proof of sufficiency is shown explicitly.
The similarity of $H$ and $H^*$ results in real eigenvalues or pairs of complex conjugate eigenvalues.
By ordering the Jordan canonical form of $H$ in real Jordan blocks and block structures of complex conjugate Jordan blocks, Zhang \textit{et al.} are able to construct the Hermitian similarity transformation $\eta$ for any Hamiltonian that is similar to its complex conjugate.
Theorem 2.1 and 2.2 are equivalent criteria for pseudo Hermiticity.

We establish a connection between pseudo Hermiticity and (anti-)unitary symmetries in non-Hermitian systems.
$\mathcal{PT}$ symmetry is defined by
\begin{equation}\label{eq:def_PTsym}
    H=\mathcal{A} H^* \mathcal{A}^{-1} \, ,
\end{equation}
with $\mathcal{A}^{-1}=\mathcal{A}^\dagger$ and $\mathcal{A}\mathcal{A}^*=1$.
A different symmetry of non-Hermitian systems is pseudo-Hermitian symmetry defined by
\begin{equation}\label{eq:def_psHsym}
    H = \varsigma H^\dagger \varsigma^{-1}\, ,
\end{equation}
where $\varsigma^{-1}=\varsigma^\dagger$ and $\varsigma^2=1$.
We emphasize the subtle difference between pseudo Hermiticity and pseudo-Hermitian symmetry.
Pseudo-Hermitian symmetry arises if we constrain the generator of pseudo-Hermiticity to be unitary. \cite{Kawabata2019,Sayyad2022,Montag2024}
Further we stress that pseudo-Hermitian symmetry is not to be confused with the non-unitary, anti-linear symmetries, which follow from pseudo Hermiticity. \cite{Mostafazadeh12002}

\textbf{Theorem 2.3.} \textit{For finite-dimensional systems, a $\mathcal{PT}$-symmetric or pseudo-Hermitian symmetric Hamiltonian H is necessarily pseudo-Hermitian.}

\textit{Proof.} By the definition of pseudo-Hermiticity and pseudo-Hermitian symmetry this is clear for the later statement.
For a $\mathcal{PT}$-symmetric system Eq.~(\ref{eq:def_PTsym}) shows that $H$ is similar to $H^*$.
Therefore, according to Theorem 2.2 the Hamiltonian is pseudo Hermitian.
This was already realized by Zhang \textit{et al.} in Ref.~\citenum{Zhang2020}.

\textbf{Theorem 2.4.} \textit{For any $H\in\mathbb{C}^{2\times 2}$, if $H$ is pseudo-Hermitian it is necessarily $\mathcal{PT}$-symmetric and pseudo-Hermitian symmetric. For finite-dimensional systems with dimension $n>2$, pseudo-Hermiticity does not imply either symmetry of the Hamiltonian.}

\textit{Proof.} $H$ has a certain symmetry if and only if there exists a unitary matrix $U$ fulfilling Eq.~(\ref{eq:def_PTsym}) or (\ref{eq:def_psHsym}) with the symmetry specific additional condition on $U$.
Note that the case $n=1$ is trivial due to the fact that $H$ reduces to a real number.
We show first whether a unitary similarity $U$ between $H$ and $H^*$ or $H^\dagger$ exists in general and then investigate the properties of $U$.
For the proof we make use of Specht's criterion. \cite{HornBook}

\textbf{Specht's criterion} \textit{The matrices $A,B\in\mathbb{C}^{n \times n}$ are unitarily similar, i.e., $A=UBU^\dagger$ with $U$ unitary, if and only if 
\begin{equation}\label{eq:Spechts_criterion}
    \tr\left[w\left(A,A^\dagger\right)\right]= \tr\left[w\left(B,B^\dagger\right)\right]
\end{equation}
for ever finite word $w$ in two letters.}

This criterion is useful, because an upper bound on the length of the words $w$ was introduced by Pearcy in Ref.~\citenum{Pearcy1962} and was later refined. \cite{Procesi1976,Formanek1991,Razmyslov1974,Laffey1986,Pappacena1997,SibirskiiBook}
For small $n$ the sets of non-redundant words one has to check is given in Ref.~\citenum{Dokovic2007}.
We make use of $n=2$ and $n=3$, where the non-redundant words $w(X,X^\dagger)$ are given by
\begin{align}
    n&=2: \quad X,\,X^2, \,XX^\dagger \, , \label{eq:specht_n2} \\
    n&=3: \quad X,\,X^2,\, XX^\dagger,\, X^3,\, X^2X^\dagger,\, X^2\left(X^\dagger\right)^2,\, X^2\left(X^\dagger\right)^2,\, XX^\dagger \, . \label{eq:specht_n3}
\end{align}
For $n=2$ we use that the traces of the three non-redundant words of $H$, $H^*$ and $H^\dagger$ are equal due to the pseudo-Hermiticity constraint.
Thus $H$ and $H^*$ as well as $H$ and $H^\dagger$ are unitarily similar.
The first condition from Eq.~(\ref{eq:def_PTsym}) for $\mathcal{PT}$ symmetry and from Eq.~(\ref{eq:def_psHsym}) for pseudo-Hermitian symmetry is therefore fulfilled.
The special properties of the unitary similarity matrices $\mathcal{A}$ and  $\eta$ can be shown by 
\begin{align}
    &H=\mathcal{A} H^* \mathcal{A}^{-1} \implies H^* =\mathcal{A}^* H \left(\mathcal{A}^{-1}\right)^* \implies \mathcal{A}\mathcal{A}^* = 1 \, , \\
    &H = \varsigma H^\dagger \varsigma^{-1} \implies H^\dagger = \varsigma H \varsigma^{-1} \implies \varsigma^2=1 \, .
\end{align}
Therefore pseudo-Hermiticity implies $\mathcal{PT}$-symmetry and pseudo-Hermitian symmetry for any $H\in\mathbb{C}^{2\times2}$.

For any $n\geq3$ to find unitary similarity it is necessary that the traces of the words for $n=3$ have to be equal, while there are more non-redundant words for $n>3$.
However, for non-normal $H$, i.e., $[H, H^\dagger] \neq 0$, equality of the word traces of $H$ and $H^*$ as well as of $H$ and $H^\dagger$ does not follow from pseudo-Hermiticity.
Thus for $n\geq3$ pseudo-Hermiticity is more general and not equivalent to $\mathcal{PT}$-symmetry or pseudo-Hermitian symmetry.

We note that normality of $H$ restores the equivalence of similarity and symmetry, which can be shown from the diagonalisability of $H$ and the pseudo-Hermitian spectral properties.
For any normal $H$ pseudo Hermiticity is equivalent to both $\mathcal{PT}$-symmetry and pseudo-Hermitian symmetry.
However, normality prohibits the emergence of EPs altogether, because the Hamiltonian must be diagonalizable in the whole parameter space.
Therefore, we consider non-normal Hamiltonians in the following for which the pseudo Hermiticity is more general then any of the two symmetries.

\textbf{Theorem 2.5.} \textit{For any $H\in\mathbb{C}^{n \times n}$, if $H$ is pseudo-Hermitian the codimension of an EP$n$ is reduced to $n-1$.}

\textit{Proof.} It has been shown in Ref.~\citenum{Sayyad2022} that the $2(n-1)$ real constraints for the emergence of an EP$n$ in the spectrum of traceless $H\in\mathbb{C}^{n \times n}$ can be cast as $\det[H] = \prod_i \epsilon_i = 0$ and $\tr [H^k] = \sum_i \epsilon_i^k = 0$ for $2\leq k<n$ with the eigenvalues $\epsilon_i$ of $H$.
The determinant and the traces are in general complex.
Pseudo Hermiticity implies the spectral symmetry $\left\{\epsilon\right\}=\left\{\epsilon^*\right\}$, which results in $\left\{\det[H], \tr[H^k]\right\}\in\mathbb{R}$.
This reduces the codimension of the EP$n$ to $n-1$.

For $\mathcal{PT}$ symmetry and pseudo-Hermitian symmetry this was shown in Ref.~\citenum{Sayyad2022}, but the symmetries are special cases of pseudo-Hermiticity according to Theorem 2.3.
From Theorem 2.4 we know that pseudo-Hermiticity is more general than the two symmetries, and it already induces the EPs in lower dimension without the need of symmetry.
Thus we have shown that not symmetry but similarity drives the emergence of exceptional points in lower dimensions.
Further the spectral structure surrounding the similarity-induced EPs is fully determined by the similarity even in the presence of the more restrictive $\mathcal{PT}$-symmetry or pseudo-Hermitian symmetry.
This spectral structure is discussed in detail in previous papers on symmetry-induced EPs. 
Symmetry-protected EP2 rings were found,\cite{Bergholtz2021} and the rich spectral features surrounding symmetry-induced EP3s, EP4s and EP5s in two dimensions have also been analyzed.
\cite{Mandal2021,Montag2024}

In addition to $\mathcal{PT}$ symmetry and pseudo-Hermitian symmetry, pseudo Hermiticity encloses two more special cases, namely Hermitian and real matrices.
We note that a Hermitian matrix is a special case of a pseudo-Hermitian symmetric systems, and a real Hamiltonian a special case of $\mathcal{PT}$-symmetric systems, with the symmetry generator being the identity operation in both cases.
Our results concerning EPs are applicable to real matrices, while Hermiticity does not allow for EPs.

\section{Pseudo anti-Hermitian systems}\label{sec:chiral}

To define pseudo anti-Hermiticity we first define skew similarity.
Two matrices $A,B\in\mathbb{C}^{n \times n}$ are said to be skew-similar to each other if there exists an invertible matrix $S$ such that
\begin{equation}\label{eq:def_skew_sim}
    A = -SBS^{-1} \, .
\end{equation}
We define pseudo anti-Hermiticity as Hermitian skew-similarity between the Hamiltonian $H$ and its adjoint $H^\dagger$.
A Hamiltonian $H$ is called pseudo anti-Hermitian if there exists an invertible Hermitian matrix $\Gamma$ such that
\begin{equation}\label{eq:def_chiral}
    H=-\Gamma H^\dagger\Gamma^{-1} \, .
\end{equation}

\textbf{Theorem 3.1.} \textit{For a matrix $H\in\mathbb{C} ^{n\times n}$, H is pseudo anti-Hermitian if and only if $\Tilde{H}:=iH$ is pseudo Hermitian.}

\textit{Proof.} To prove this Theorem insert $\Tilde{H} = iH$ in the definition of pseudo Hermiticity Eq.~(\ref{eq:def_psH}) and take $\Gamma=\eta$ as the similarity matrix.

Due to this relation between pseudo-Hermitian and pseudo anti-Hermitian matrices we can straightforwardly deduct properties of pseudo anti-Hermitian matrices.

\textbf{Corollary 3.1.} \textit{The eigenvalues $\epsilon$ of pseudo anti-Hermitian matrices fulfill $\left\{\epsilon\right\}=\left\{-\epsilon^*\right\}$.}

\textbf{Corollary 3.2.} \textit{For a matrix $H\in\mathbb{C} ^{n\times n}$, H is pseudo anti-Hermitian if and only if it is skew-similar to its complex conjugate $H^*$.}

A proof of Corollary 3.2 by explicit construction of the similarity generator is carried out analogous to the construction of the generator of pseudo-Hermiticity in Ref.~\citenum{Zhang2020}.

We now show the connection between pseudo anti-Hermiticity and anti-$\mathcal{PT}$ symmetry and chiral symmetry.
We define anti-$\mathcal{PT}$ symmetry by
\begin{equation}\label{eq:def_CPsym}
    H=-\Theta H^* \Theta^{-1} \, ,
\end{equation}
where $\Theta^{-1}=\Theta^\dagger$ and $\Theta\Theta^*=1$.
From this definition it is clear that a matrix $H$ is anti-$\mathcal{PT}$-symmetric if and only if the matrix $\Tilde{H}:=iH$ is $\mathcal{PT}$-symmetric.
We note that anti-$\mathcal{PT}$ symmetry is sometimes referred to as the combination of particle-hole symmetry and parity due to the notation of particle-hole symmetry in Hermitian systems. \cite{Kawabata2019}
However, particle-hole symmetry is defined via the transpose of a Hamiltonian and not the complex conjugation. \cite{Kawabata2019,Lieu2018}
In the Hermitian case we find $H^T=H^*$ thus particle-hole symmetry is often described via complex conjugation, which leads to the ambiguous nomenclature.
Further chiral symmetry is defined as
\begin{equation}\label{eq:def_CS}
    H=-\gamma H^\dagger \gamma^{-1} \, ,
\end{equation}
with $\gamma^{-1}=\gamma^\dagger$ and $\gamma^2=1$.
The subtle difference in the properties of the similarity matrix between pseudo anti-Hermiticity and chiral symmetry is emphasised here.
A matrix $H$ is chirally symmetric if and only if $\Tilde{H}:=iH$ has pseudo-Hermitian symmetry.
The name chiral symmetry follows from the Altland-Zirbauer classification, where the combination of time-reversal symmetry and particle-hole symmetry is named chiral symmetry. \cite{Altland1997}
Considering the non-Hermitian symmetries this results in Eq.~\eqref{eq:def_CS}. \cite{Kawabata2019}

Due to the connection between the $\mathcal{PT}$ and the pseudo-Hermitian symmetry of $\Tilde{H}:=iH$ to the anti-$\mathcal{PT}$ and the chiral symmetry of $H$, respectively, we can infer properties of pseudo anti-Hermitian matrices from the theorems 2.3, 2.4 and 3.1.

\textbf{Corollary 3.3.} \textit{For finite-dimensional systems, an anti-$\mathcal{PT}$-symmetric or chiral-symmetric Hamiltonian H is necessarily pseudo anti-Hermitian.}

\textbf{Corollary 3.4.} \textit{For any $H\in\mathbb{C}^{n\times n}$, if $n=2$ and $H$ is pseudo anti-Hermitian it is necessarily anti-$\mathcal{PT}$-symmetric and chiral symmetric. For finite-dimensional systems with dimension $n>2$, pseudo anti-Hermiticity does not imply either symmetry of the Hamiltonian.}

The equivalence of pseudo anti-Hermiticity and anti-$\mathcal{PT}$-symmetry as well as chiral symmetry can be restored by enforcing normality on the Hamiltonian $H$.
However, this would disallow exceptional points to emerge in the systems as already mentioned before.

\textbf{Corollary 3.5.} \textit{For any $H\in\mathbb{C}^{n \times n}$, if $H$ is pseudo anti-Hermitian the codimension of an EP$n$ is reduced to $n-1$.}

While this Corollary follows from Theorem 2.5 and 3.1 we expand on this here in more detail to clarify the reasoning for the pseudo anti-Hermitian system.
We consider the complex conditions $\det [H]=0$ and $\tr [H^k]=0$ for $2\leq k<n$.
The spectral symmetry $\left\{\epsilon\right\}=\left\{-\epsilon^*\right\}$, which is a consequence of the pseudo anti-Hermiticity, reduces the number of constraints, because the determinant of $H$ and each of the traces of $H^k$ is either real or purely imaginary.
This reduces the codimension of the EP to $n-1$.
For the two symmetries enclosed by pseudo anti-Hermiticity according to Corollary 3.3 this was shown in Ref.~\citenum{Sayyad2022}.
However, from Corollary 3.4 it is clear that pseudo anti-Hermiticity is more general than either anti-$\mathcal{PT}$-symmetry or chiral symmetry.
Pseudo anti-Hermiticity already induced EPs by lowering their codimension without the need of symmetries of the Hamiltonian.
Further, the spectral structure surrounding a pseudo anti-Hermiticity-induced exceptional point is fully determined by the skew-similarity of the Hamiltonian.
For EP3s in two dimensions the spectral structure is equivalent to the structures described in Ref.~\citenum{Montag2024} for anti-$\mathcal{PT}$-symmetry and chiral symmetry.

Besides anti-$\mathcal{PT}$-symmetric and chiral-symmetric systems there are two notable special cases of pseudo anti-Hermitian systems.
The first case is anti-Hermiticity of $H$, meaning $H=-H^\dagger$, which can be interpreted as chiral symmetry with the identity as generator.
Because anti-Hermiticity implies normality no exceptional points can emerge in anti-Hermitian systems.
The other case are imaginary matrices $H=-H^*$, which are anti-$\mathcal{PT}$-symmetric with the identity as generator. 
Our results are thus also applicable for imaginary matrices.

\section{Self skew-similar systems}\label{sec:skew}

With skew-similarity defined in Eq.~(\ref{eq:def_skew_sim}) any Hamiltonian $H$ is self skew-similar if it anticommutes with an invertible matrix $S$.
The self skew-similarity follows from this definition by rearranging
\begin{equation}\label{eq:def_self_skew}
    \{H,S\}=HS+SH=0 \iff H=-SHS^{-1} \, .
\end{equation}

The self skew-similarity constraints the spectrum to $\left\{\epsilon\right\}=\left\{-\epsilon\right\}$.
This can be shown by considering an eigenstate $\ket{\chi}$ of the self skew-similar Hamiltonian $H$ with eigenvalue $\epsilon$.
Applying the definition Eq.~(\ref{eq:def_self_skew}) yields
\begin{equation}
    -S^{-1}H\ket{\chi} = -\epsilon\left(S^{-1}\ket{\chi}\right) = H \left(S^{-1}\ket{\chi}\right) \, ,
\end{equation}
and this shows the spectral symmetry.

The two symmetries that enforce the same spectral constraint on the system are sublattice symmetry and $\mathcal{CP}$-symmetry.
Sublattice symmetry is defined by 
\begin{equation}
    H=-\mathcal{S}H\mathcal{S}^{-1} \, ,
\end{equation}
with $\mathcal{S}^{-1}=\mathcal{S}^\dagger$ and $\mathcal{S}^2=1$.
Again we emphasize the subtle difference in the properties of the generators of self skew-similarity and the unitary sublattice symmetry.
For Hermitian systems sublattice symmetry is equivalent to chiral symmetry, however, the non-Hermiticity results in the distinction of the two symmetries, which have different properties for non-Hermitian systems.\cite{Kawabata2019,Sayyad2022,Montag2024}
We define $\mathcal{CP}$-symmetry as
\begin{equation}
    H=-XH^TX^{-1} \, ,
\end{equation}
where $X^{-1}=X^\dagger$ and $XX^*=1$.
This symmetry is the combination of particle-hole symmetry and parity for non-Hermitian systems, where particle-hole symmetry is defined using the transpose of the matrix $H$. \cite{Kawabata2019}
In the literature this is sometimes referred to as pseudo-chiral symmetry, but this nomenclature is misleading, because the symmetry is not related to chiral symmetry.

\textbf{Theorem 4.1.} \textit{For finite-dimensional systems, a sublattice-symmetric or $\mathcal{CP}$-symmetric Hamiltonian $H$ is necessarily self skew-similar.}

\textit{Proof.} By the definition of sublattice symmetry it is clear that a sublattice-symmetric Hamiltonian is self skew-similar.
Because every matrix is similar to its transpose, $\mathcal{CP}$-symmetry entails self skew-similarity.

\textbf{Theorem 4.2.} \textit{For any Hamiltonian $H\in\mathbb{C}^{n\times n}$, self skew-similarity does not imply either symmetry of the Hamiltonian for $n\geq2$.}

\textit{Proof.} For any $H\in\mathbb{C}^{n\times n}$ with $n\geq3$ Specht's criterion is not fulfilled for a generic self skew-similar matrix.
For $H\in\mathbb{C}^{2\times 2}$ Specht's criterion is always fulfilled, however, the additional properties enforced on the unitary operator to be a symmetry generator are not fulfilled for either sublattice symmetry or $\mathcal{CP}$-symmetry.
Note that $n=1$ is a special case, because the self skew-similarity implies $H=0$, which has arbitrary unitary and anti-unitary symmetries.

\textbf{Theorem 4.3.} \textit{For any $H\in\mathbb{C}^{n \times n}$, if $H$ is self skew-similar the codimension of an EP$n$ is reduced to $n$ if $n$ is even and to $n-1$ if $n$ is odd.}

\textit{Proof.} We consider the complex conditions $\det [H]=0$ and $\tr [H^k]=0$ for $2\leq k<n$.
For any odd $k$ the trace $\tr [H^k]$ vanishes for any self skew-similar Hamiltonian due to the spectral constraint $\left\{\epsilon\right\}=\left\{-\epsilon\right\}$.
For odd $n$ the determinant always vanishes, because
\begin{equation}
    \det [H] = \det [S]\det [-H]\det [S^{-1}] = (-1)^n \det [H] \overset{n\in\textrm{odd}}{\implies} \det [H] = 0 \, .
\end{equation}
This reduces the codimension in the case of odd $n$ to $n-1$ and for even $n$ the codimension of EP$n$s is reduced only to $n$.

For the two symmetries that realize self skew-similar Hamiltonians this was shown in Ref.~\citenum{Sayyad2022}.
From Theorem 3.3 it is clear that self skew-similarity is more general than either of the two symmetries.
According to Theorem 3.4 self skew-similarity already induces EP$n$s by lowering their codimension.
Because the spectral symmetry is enforced by the similarity, symmetry of the Hamiltonian is not needed.
Further the spectral structure accompanying the similarity-induced EPs is determined by the self skew-similarity, and also not affected by the additional constraints of $\mathcal{CP}$-symmetry or sublattice symmetry.

A special case of self skew-similarity are anti-symmetric Hamiltonians $H=-H^T$, which are $\mathcal{CP}$-symmetric with the identity as generator.
All our results are applicable for anti-symmetric matrices.

\section{Conclusions}\label{sec:conclusion}

It was previously shown that $\mathcal{PT}$-symmetry entails pseudo Hermiticity for finite dimensional systems.
In this paper we show that this relation can be generalized to all unitary and anti-unitary symmetries, which lower the codimension of exceptional points.
We proof that each of these symmetries is a special case of one of three generalized similarities, namely pseudo Hermiticity, pseudo anti-Hermiticity and self skew-similarity.
Each similarity encompasses two symmetries, and in the case of pseudo Hermiticity and pseudo anti-Hermiticity for finite-dimensional systems of size $n>2$ the similarities are more general than the respective symmetries.
In the case of self skew-similar Hamiltonians this even holds for $n\geq2$.

Overall we find that the spectral features of non-Hermitian systems and the emergence of stable EPs is linked to the relevant similarity, and not the symmetry of the system contrary to previous assumptions.
The similarities are far less restrictive compared to unitary or anti-unitary symmetries.
As such, the presence of similarities may lead to the robustness of symmetry-stabilized non-Hermitian features to symmetry-breaking perturbations.

\begin{acknowledgments}
We are grateful to Julius Gohsrich and Alexander Felski for insightful discussions. We acknowledge funding from the Max Planck Society Lise Meitner Excellence Program 2.0.
We also acknowledge funding from the European Union via the ERC Starting Grant “NTopQuant”. 
Views and opinions expressed are however those of the authors only and do not necessarily reflect those of the European Union or the European Research Council (ERC). Neither the European Union nor the granting authority can be held responsible for them.
\end{acknowledgments}

\bibliography{references_symmetry.bib}

\end{document}